\documentclass[preprint,showpacs,preprintnumbers,amsmath,amssymb]{revtex4}

\usepackage{graphicx}
\usepackage{dcolumn}
\usepackage{bm}
\usepackage{color}

\newcommand{\irm}{\mathrm{i}}

\begin{document}

\title{Experimental demonstration of a Displacement noise Free Interferometry scheme
for gravitational wave detectors showing displacement noise reduction at low frequencies}

\author{Antonio Perreca$^1$}
 \email{perreca@star.sr.bham.ac.uk}
\author{Simon Chelkowski$^1$}
\author{Stefan Hild$^2$}
\author{Andreas Freise$^1$}%
\affiliation{%
$^{1}$School of Physics and Astronomy, University of Birmingham,\\ Edgbaston, Birmingham, B15 2TT, United Kingdom\\
$^{2}$Institute for Gravitational Research, University of Glasgow, Glasgow, G12 8QQ, United Kingdom}%

\date{\today}

\begin{abstract}
This paper reports an experimental demonstration of partial displacement
noise free laser interferometry in the gravitational wave
detection band. The used detuned Fabry-Perot cavity allows
the isolation of the mimicked gravitational wave signal from the
displacement noise on the cavities input mirror. By properly
combining the reflected and transmitted signals from the cavity a
reduction of the displacement noise was achieved. Our results
represent the first experimental demonstration of this recently
proposed displacement noise free laser interferometry scheme.
Overall we show that the rejection ratio of the displacement noise
to the gravitational wave signal was improved in the frequency range
of 10\,Hz to 10\,kHz with a typical factor of $\sim$\,60.
\end{abstract}


\maketitle

\section{\label{sec:Intro}Introduction to Displacement
noise Free Interferometry}

General Relativity describes gravity as the curvature of space-time.
The theory predicts the existence of \textit{gravitational waves} (GWs)
which can be described as ripples of space-time propagating with the
speed of light.

The detection of GWs is possible by measuring the variation
$\delta l$ of the distance between two free masses but the
predicted GW amplitude is so small which renders the detection very
difficult. So far no instrument has detected any GW signals
directly.

The sensitivity of the current GW detectors is limited by several
noise sources. One group, usually referred to as
\textit{displacement noises} (DNs) directly moves the reflective part of
the test masses. Current GW detectors are limited by displacement
noise such as seismic noise, gravity gradient noise, thermal noise
and radiation pressure noise at frequencies below $100\,$Hz.

The technology development for GW detectors has focused on
reducing each of the noise contributions independently, e.g.
suspension systems are employed to decouple optical components
from the seismic motion of the environment.
Several new ideas and concepts are under study to create a new generation
of GW detectors with a strongly improved sensitivity
\cite{Freise09,Edgar09,Vahlbruch05,Chen09}. In the context of future GW detectors a new idea
called displacement and frequency noise free interferometry (DFI)
was proposed by
S.Kawamura and Y.Chen \cite{Kawamura04}. DFI is based on the fact
that gravitational waves and displacement noise as well as
frequency noise affect the light in a different manner and aims at
reducing all displacement noises and frequency noise
simultaneously. The realization of an experiment with multiple
read-out channels where each single channel carries the
gravitational wave signal and the noise information differently
allows the creation of a channel
that is completely free from frequency and displacement noises \cite{Chen06a,Chen06b}.

The current experimental demonstrations of DFI affected only
frequencies above $\sim1\,$MHz \cite{Sato07, kokeyama09}.
However, the large-baseline gravitational wave detectors do not
work in this band. Recently a new DFI scheme has been proposed
which works in a low frequency region. A detuned Fabry-Perot (FP)
cavity configuration \cite{TV08} in combination with two lasers is
used to partially remove the displacement noise from both cavity
mirrors. One laser is used for the input cavity mirror (IM) and
one is used for the end cavity mirror (EM) (Double Pumped
Fabry-Perot cavity). Such a configuration, although it does not
include the frequency noise aspect of DFI, allows the isolation of
GW signal from displacement noise in a wide range of frequencies.
Basically for each laser the reflected and transmitted output
signals of the detuned FP-cavity carry different GW and
displacement noise information, due to the existence of the prompt
reflected light. A proper combination of both signals results
in the suppression of the displacement noise of the cavity's input
mirror. Here the mechanism of noise cancelation is completely
different from the Chen-Kawamura's mechanism. The latter uses the
distributed nature of GW's which results in different kind of
responses. In the long wave approximation ($\lambda_{gw}\gg L$),
where $\lambda_{gw}$ is the GW wavelength and L is the cavity
length, the leading order of the DFI signal for the detuned
FP-cavity is $h(L/\lambda_{gw})^0$ which is much better than the
$h(L/\lambda_{gw})^2$ that can be obtained from Chen-Kawamura DFI
scheme \cite{Chen06b}. Nevertheless the detuned FP-cavity scheme
looses the optical resonant gain from the cavity which is in the
order of $c/\gamma L$, where $\gamma$ is the cavity half
bandwidth. Hence, the sensitivity of this scheme concerning GWs is
strongly reduced compared to conventional interferometers and the
noise performance of auxiliary optics becomes much more important.
Hence, our experiment shows partial DFI because it is not
completely DN free. Nevertheless we stick to the name DFI
throughout this paper to be compatible with previous published
papers \cite{TV08} and address as DFI only the suppression of
displacement noise of the cavity mirrors.

In this letter we present the first experimental proof of
principle demonstration of the detuned FP cavity based DFI scheme
proposed in \cite{TV08}.
We could thus verify the core concept of this new idea
which is the basis for new proposed interferometer schemes \cite{Rakhubovsky08}.
We use one
laser in combination with two homodyne detectors to strongly
suppress the displacement noise of the input mirror of a FP-cavity
with respect to a simulated GW signal. As a result we
gain a factor of $\sim\,$60 in the GW signal to displacement noise
ratio in the whole frequency range of interest. 

\section{DFI configuration: Detuned FP-cavity}
\begin{figure}[t]
\includegraphics[width=8.6cm,keepaspectratio]{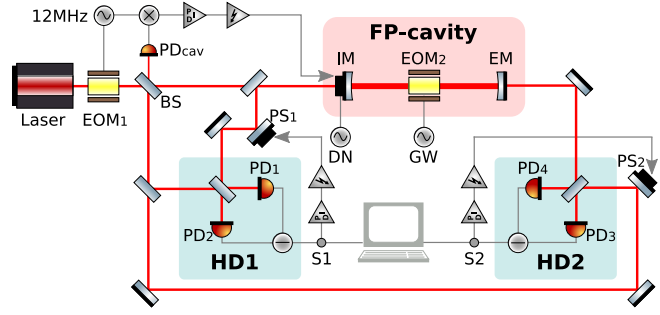}
\caption{Experimental setup of our DFI experiment using a detuned
FP cavity. The laser pumps the cavity through the mirror IM. The
reflected light is measured with homodyne detector HD1 while the
transmitted light is measured with homodyne detector HD2. The cavity length is
30 cm, the bandwidth $2\gamma$ is $2.4\,$MHz and the detuning
$\delta$ is $12\,$MHz. The cavity is controlled in its detuned state
with the Pound-Drever-Hall technique: The feedback signal is
applied to the PZT attached to the input cavity mirror. The two
homodyne detectors are controlled using the difference
photocurrents $S1$ and $S2$ as error signal while the feedback is
applied to the phase shifters PS1 and PS2 respectively. $S_1$ and $S_2$
are used for the transfer function measurements of the
displacement noise and GW signal as well.} \label{fig:figure1}
\end{figure}

A detuned FP-cavity pumped through one side (see Fig.
\ref{fig:figure1}) guarantees the existence of one channel in
reflection (S1) and one in transmission (S2) which contain GW and
DN information in a different ratio. The difference is due to the
existence of the directly reflected light which occurs only on the
input cavity mirror. This directly reflected light contains only
the information about the position of mirror IM and not
the position of mirror EM \cite{TV08}.
The response of the transmitted signal measured in S2 can be written as:
\begin{eqnarray}
S2=q_1(\phi_\mathrm{GW}+\phi_\mathrm{EM}-\phi_\mathrm{IM})+\phi_\mathrm{S2},
\label{eq:one}
\end{eqnarray}
where $q_1$ represent the
resonant gain factor of the cavity, $\phi_\mathrm{GW},\,\phi_\mathrm{EM},\,\phi_\mathrm{IM}$
are the phases
accumulated in the cavity due respectively to GW signal and the
displacements of both the cavity mirrors EM and IM
and $\phi_\mathrm{S2}$ is the phase
induced by the optical elements which the light encounters before
it is detected in $S2$.
The response
of the reflected signal measured in S1 is written:
\begin{eqnarray}
S1=p(\phi_\mathrm{IM}-\phi_\mathrm{S1})+q_2(\phi_\mathrm{GW}+\phi_\mathrm{EM}-\phi_\mathrm{IM}),
\label{eq:two}
\end{eqnarray}
where $p$ describes the 'prompt' reflected light from the
input cavity mirror, $q_2$ is the resonant gain factor of the
cavity, and $\phi_\mathrm{S1}$ describes the phase changes induced by all
the auxiliary optical elements the light passes before it is
detected in S1. Using only one laser one can find a certain linear
combination of the reflected and the transmitted signals which
will partially remove the displacement noise fluctuation from
mirror IM while the displacement noise of mirror EM and the
simulated GW information remain. If a second laser is used and
coupled into the cavity through mirror EM simultaneously and
another two homodyne detectors are set up, two more output
channels are available. Only a proper linear combination S of all
four output channels allows to suppress the displacement noise of
both mirrors while the GW signal is retained. Using the
approximation $\delta\tau$, $\gamma\tau\ll 1$ with $\tau=L/c$, the
DFI response in the latter case for a cavity with two equal
mirrors and $L$ length is given by \cite{TV08}:
\begin{eqnarray}
S=p(\phi_\mathrm{GW}+\phi_\mathrm{S2}-\phi_\mathrm{S1})\approx
\frac{\irm\delta}{\gamma-\irm\delta}(\phi_\mathrm{GW}+\phi_\mathrm{S2}-\phi_\mathrm{S1}),
\label{eq:three}
\end{eqnarray}
where $\delta$ and $\gamma$ are respectively the cavity detuning
and the cavity half bandwidth. It can be noticed that the optimal case of
the GW response is given when the $p$ factor is approximately one
which requires a large
detuning compared with the cavity half bandwidth ($\delta\gg\gamma$).

\section{Experimental set-up}\label{sec:level2}
Our experimental setup is shown in Fig. \ref{fig:figure1}. The
laser source is a commercial solid state Nd:YAG yielding $1\,$W at
1064\,nm. The light originating from the laser is split into two
beams, one to pump the FP cavity and one to provide the two local
oscillators (LO) for the homodyne detectors HD1 and HD2. The FP
cavity is formed by two identical mirrors which are separated
by 30\,cm. Each mirror has a power reflectivity of 98.5\,\% and a
radius of curvature of 1\,m resulting in a cavity bandwidth $2\gamma$ of
2.4\,MHz.

$\mathrm{PD_{cav}}$ is used to detect the reflected light from the
cavity and to generate an error signal using the Pound-Drever-Hall
(PDH) technique. The electro-optic-modulator EOM1 is used to
imprint phase modulation sidebands with a frequency of $12\,$MHz
onto the laser light. The photocurrent produced by
$\mathrm{PD_{cav}}$ is then demodulated with the same frequency
and filtered to generate a PDH like error signal. This error
signal is processed and fed back to a PZT being attached to the
input cavity mirror to stabilize the cavity in a detuned state,
shifted by $12\,$MHz from the cavity's resonance.

The reflected and transmitted signals from the cavity are
individually sensed with the two homodyne detectors HD1 and HD2. These 
allow us to measure signals in an arbitrary
quadrature in between amplitude and phase quadrature. We used a
local oscillator power of $25\,$mW for each homodyne detector. The
power of the reflected signal beam at the cavity is $2\,$mW whereas
the transmitted signal beam through the cavity signal power
is $0.1\,$mW. Thereby we fulfilled the condition that the LO power
has to be much higher than the signal power to ensure that the
resulting signal is dominated by the signal on the signal beam and
not by noise present on the LO \cite{Bachor98}.

The difference photocurrent S1 of homodyne detector HD1 as well as
the difference photocurrent S2 of homodyne detector HD2 are used
to generate individual error signals for the homodyne detectors.
Each error signal is fed back to the PZT actuators PS1 and PS2
respectively. Thereby we provide the necessary control to lock
both homodyne detectors to phase quadrature. The control bandwidth
of the cavity and the homodyne detector control loops are kept as
low as possible, around $\sim70\,$Hz, in order to avoid that the
control loop affects the DFI response in the low frequency region.

The electro-optic-modulator EOM2 is used to imprint a phase
modulation on the light resonating inside the cavity, as would be
done by a GW. Hence EOM2 is used to create our simulated GW
signal. By injecting a swept-sine signal into EOM2, we can measure
a simulated GW transfer function to both homodyne detectors.
The original scheme proposed in \cite{TV08} includes the
effects of the GW on the LO paths. Our scheme however represents
the case where the LO for HD2 can be provided by an independent laser.
The DN rejection ratio between these two schemes can
differ at maximum by a factor of two.
Hence our experiment shows qualitatively the proof of principle of the
originally proposed scheme.

The
PZT attached to mirror IM stabilizes the cavity length by applying
a feedback signal. Furthermore, our simulated displacement noise
signal is imprinted on the light by applying swept-sine signal to
this PZT which allows us to measure a displacement noise transfer
function to both homodyne detectors. Both transfer functions (DN
and simulated GW) are measured using the homodyne outputs S1 and
S2, thereby creating the basis for the demonstration of this DFI
scheme.

\begin{figure}[t!]
\includegraphics[width=8.6cm,keepaspectratio]{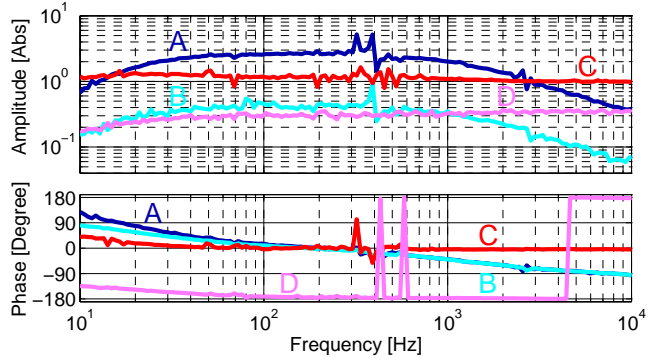}
\caption{Comparison of the measured transfer functions with the
two homodyne detectors HD1 and HD2. The two displacement noise
transfer functions are represented by trace A and B, whereas trace C
and D show the transfer functions for the simulated GW signals.
One can see that the two displacement noise transfer functions are
in phase almost in the whole frequency range, where the simulated
GW signal transfer functions have a relative phase shift of
$180^\circ$ to each other.}
\label{fig:figure2}
\end{figure}

\section{Measurements and results}

To demonstrate the detuned FP cavity based DFI scheme, we measured
the displacement noise and simulated-GW responses. The resulting
transfer functions are shown in Fig.~\ref{fig:figure2}. Both, the
signal as well as the noise strength measured in the homodyne
detectors depend on the particular quadrature used. Hence, it is
important that the quadrature control is stable.
In particular we ensured that all of our data obtained with
the two homodyne detectors were measured in the same quadrature,
namely the phase quadrature.

\begin{figure}[t!]
\includegraphics[width=8.6cm,keepaspectratio]{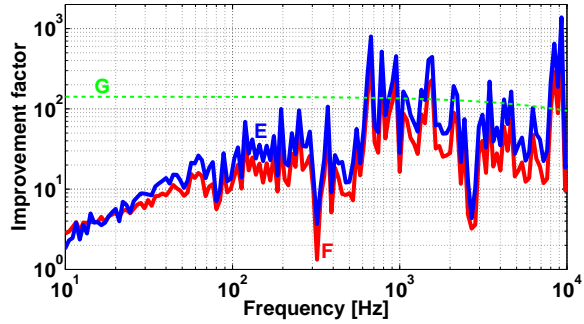}
\caption{The improvement due to the DFI scheme is expressed by the
ratio $\rho_i$ between the processed DN rejection factor
$\sigma_\mathrm{DFI}$ and unprocessed DN rejection factors
$\sigma_\mathrm{S1}$ and $\sigma_\mathrm{S2}$. The two resulting
ratios $\rho_\mathrm{S1} = \sigma_\mathrm{DFI}/\sigma_\mathrm{S1}$
(trace E) and $\rho_
\mathrm{S2}=\sigma_\mathrm{DFI}/\sigma_\mathrm{S2}$ (trace F)
show an improvement of $\sim$\,60 in the whole
frequency range with a slight advantage for $\rho_\mathrm{S1}$
compared to $\rho_\mathrm{S2}$. Trace G shows the expected result
for a phase difference between the transfer functions of the two
homodyne detectors of $\phi=1^\circ$, which reduces the maximally
achievable DN reduction factor to $\sim$\,140.}\label{fig:figure3}
\end{figure}

As one can see the two transfer functions for the displacement
noise from the cavity input mirror IM to the two homodyne
detectors $S1_\mathrm{DN}$ and $S2_\mathrm{DN}$ (trace A and B in
Fig.~\ref{fig:figure2}) are in phase in almost the entire
frequency range. On the other hand the two GW signal transfer
functions $S1_\mathrm{GW}$ and $S2_\mathrm{GW}$ (trace C and D in
Fig.~\ref{fig:figure2}) are out of phase by about $180^\circ$. All
transfer functions include phase shifts induced by the optical
elements ($\phi_\mathrm{S1},\phi_\mathrm{S2}$) that the reflected and
transmitted light from the cavity passes before the detection in the
homodyne detectors $S1$ and $S2$ respectively. The decreasing
magnitude of all transfer functions towards low frequencies is a
result of the cavity servo loop gain which increases at low
frequencies thereby suppressing the injected signals more
strongly.

The fact that the GW and DN transfer functions have different
phase relations can be used to create two new DFI data channels
where the GW content is maintained while the DN content will be
strongly suppressed. These two new data channels are given by:
\begin{eqnarray}
S_\mathrm{DN,DFI}=S1_\mathrm{DN}-k\cdot S2_\mathrm{DN},
\\
S_\mathrm{GW,DFI}=S1_\mathrm{GW}-k\cdot S2_\mathrm{GW}.
\label{eq:fourandfive}
\end{eqnarray}
here $k$ represents a fixed scaling factor which minimizes the DN
content in channel $S_\mathrm{DN,DFI}$. In our case we arbitrarily
choose $k$ to be the ratio of the DN transfer function magnitude at
50\,Hz
($k=S1_\mathrm{DN_{A}}\![50\,\mathrm{Hz}]/S2_\mathrm{DN_{A}}\![50\,\mathrm{Hz}]$),
as changing the frequency for determining $k$ does not
dramatically change the results.

The DN rejection factor $\sigma$ of the initial
unprocessed data channels of $S1$ and $S2$ are given by
$\sigma_\mathrm{S1}=S1_\mathrm{GW}/S1_\mathrm{DN}$ and
$\sigma_\mathrm{S2}=S2_\mathrm{GW}/S2_\mathrm{DN}$ and show us how
good a GW can be detected with respect to the present DN. For the
processed data channels a similar DN rejection factor given by
$\sigma_\mathrm{DFI}=S_\mathrm{GW,DFI}/S_\mathrm{DN,DFI}$ can be
calculated.  To see the enhancement effect of the DFI in our
experiment we plot in Figure~\ref{fig:figure3} the ratio $\rho_i$
between the processed and unprocessed DN rejection factorss
$\rho_\mathrm{S1} = \sigma_\mathrm{DFI}/\sigma_\mathrm{S1}$ (trace A) and $\rho_
\mathrm{S2}=\sigma_\mathrm{DFI}/\sigma_\mathrm{S2}$ (trace B)
respectively.

The reduction of the DN shown by $\rho_i$ is significant in the
frequency range of interest.
Overall $\rho_\mathrm{S1}$ performs a little bit better
than $\rho_\mathrm{S2}$.
In the whole frequency range of interest the DN is reduced by a
factor of $\sim$\,60.

The DFI toy model described in \cite{TV08} provides a perfect 
cancelation of displacement noise from the input cavity mirror,
which corresponds to an infinite DN rejection. A theoretical
description of our experiment using ideal components predicts
perfect cancelation only at DC with a 1/f frequency dependence.
However, realistic rejection ratios must be computed including
inevitable asymmetries in the experimental setup.
If the transfer functions in Fig.~\ref{fig:figure2}
(i.e. the DN transfer functions marked with A and B) show a phase
difference $\phi$ the expected improvement factor can be
expressed as $\alpha /(e^{i\phi}-1)$
where $\alpha=|S_\mathrm{GW,DFI}|/|S1_\mathrm{GW}|$.
In particular when $\phi=0.1^\circ$ the expected improvement factor is
$\sim$\,1500.
Whereas a phase difference of $\phi=1^\circ$ reduces the
improvement factor to $\sim$\,140 resulting in trace G shown in
Fig.~\ref{fig:figure3}. As one can see the overall DN reduction
level of trace G corresponds quite well with our experimental
result at high frequencies.
In addition to this frequency independent phase difference, which
we expect from an imperfect setup of the homodyne detectors, we
could also identify a frequency dependent asymmetry. This
originates from slight differences in the feedback control
electronics and lead to different slopes in the phase behaviour at
low frequencies. In more detail, traces A and B of
Fig.~\ref{fig:figure2} have relatively high phase difference at
low frequencies which decreases up to $\sim100\,$Hz while the
corresponding amplitudes have flat shapes starting from
$\sim70\,$Hz. Less dominant but still present, this effect is
visible in traces C and D. Due to this frequency dependent phase
difference the resulting DN rejection factor is decreasing towards
low frequencies and does not follow the expected behaviour shown
by trace G.
Furthermore, this type of table-top experiment is subject to
mechanical vibrations of optics mounts which create sharp
dispersion-like structures in the DN and GW transfer functions at
frequencies between 200\,Hz and 4\,kHz. The phase asymmetries
mentioned above convert such dispersion structures in peaks or
dips in the DN rejection factor.

\section{Conclusions}

In conclusion, we have demonstrated the first experimental proof of
principle of the detuned FP cavity based DFI scheme showing a
large enhancement of a mimicked GW signal compared to the DN in the
gravitational frequency band from 10\,Hz-10\,kHz. In particular we
used a symmetrical and detuned FP cavity in combination with two
homodyne detectors to create two data channels each containing
information about the simulated GW signal and the DN. We processed
the data of these two channels and created one new DFI channel in
which the DN of the IM of the FP cavity was strongly suppressed. A
detailed analysis of the performance improvement within the GW
frequency band showed that at all frequencies the GW signal to DN
ratio was improved with a typical factor of $\sim$\,60.
Although these results are promising the main problem of this
detuned FP cavity DFI scheme is that the enhancement from the
cavity effect is lost. Hence, the displacement noise of any
auxiliary optics becomes more important.
{Commonly high finesse cavities are used in conventional GW
interferometers to enhance the GW signal size and therefore minimize
the influence of the displacement noises of auxiliary optics.
The demonstrated DFI scheme however uses a different approach
where the cavity finesse is suppressed together with the displacement
noise of the cavity mirrors. This can be beneficial to the conventional
method if relative displacement of nearby optics is relatively small.
A possible solution is
presented in \cite{Rakhubovsky08} where two double
pumped cavities with mirrors attached to rigid platforms are described.
This idea is currently under
investigation.

\section{Acknowledgments}
We would like to thank S.P.~Tarabrin and S.P.~Vyatchanin for their
patience in explaining their DFI scheme to us and for the many
fruitful discussion that followed. This work has been supported by
the Science and Technology Facilities Council (STFC) and the
European Gravitational Observatory (EGO). This document has been
assigned the LIGO Laboratory document number LIGO-P0900292

\bibliography{bham-ifolab}

\end{document}